\documentstyle[prl,aps,floats,a4,epsf]{revtex}

\preprint{DAMTP-1999-183}
\date{\today}
\tighten
\begin{document}
\draft
\def\sqr#1#2{{\vcenter{\hrule height.3pt
       \hbox{\vrule width.3pt height#2pt  \kern#1pt
          \vrule width.3pt}  \hrule height.3pt}}}
\def\square{\mathchoice{\sqr67\,}{\sqr67\,}\sqr{3}{3.5}\sqr{3}{3.5}}


\title{Can our Universe be inhomogeneous on large sub-horizon scales?}

\author{P.\ P.\ Avelino${}^{1,2}$\thanks{
Electronic address: pedro\,@\,astro.up.pt}
J.\ P.\ M.\ de Carvalho${}^{1,3}$\thanks{
Electronic address: mauricio\,@\,astro.up.pt}
C.\ J.\ A.\ P.\ Martins${}^{4,1}$\thanks{
Electronic address: C.J.A.P.Martins\,@\,damtp.cam.ac.uk}
and J.\ C.\ R.\ E.\ Oliveira$^{1}$}

\address{${}^1$ Centro de Astrof\'{\i}sica, Universidade do Porto\\
Rua das Estrelas s/n, 4150-762 Porto, Portugal}

\address{${}^2$ Dep. de F{\' \i}sica da Faculdade de Ci\^encias da
Univ. do Porto\\
Rua do Campo Alegre 687 -- 4169-007 Porto -- Portugal}

\address{${}^3$ Dep. de Matem\'atica Aplicada da Faculdade de Ci\^encias da
Univ. do Porto\\
Rua das Taipas, 135 -- 4050 Porto -- Portugal}

\address{${}^4$ Department of Applied Mathematics and Theoretical Physics\\
Centre for Mathematical Sciences, University of Cambridge\\
Wilberforce Road, Cambridge CB3 0WA, U.K.}

\maketitle
\begin{abstract}

{We show that our Universe may be inhomogeneous on large sub-horizon
scales  without us being able to realise it.
  We assume that a network of domain walls permeates the  universe dividing
it in domains with slightly different vacuum energy densities.   We
require that the energy scale of the phase transition which produced the
domain walls is sufficiently low so that the walls  have a negligible
effect on structure formation. Nevertheless,  the different vacuum
densities of  different domains will lead to different values of the
cosmological  parameters $\Omega_\Lambda^0$, $\Omega_m^0$ and $h$, in each
patch  thus affecting the growth of cosmological perturbations at recent
times.
  Hence, if our local patch of the universe (with uniform vacuum density) is
big enough -- which is likely to happen given that we should have on
average about one domain per horizon volume -- we might not notice these
large-scale inhomogeneities. This happens because in order to see a patch
with a different vacuum density one may have to look back at a time when
the universe was still very homogeneous. }

\end{abstract}
\newpage

\section{Introduction}
\label{secintro}

The last year or so has seen the first serious attempts to provide some direct
connections between  ``fundamental'' high-energy physics
\cite{Polchinski} and ``low-energy'' standard cosmology
\cite{Kolb}.
  Although this ``top-down'' approach is still at a very early stage, 
a number of
crucial general trends already became apparent.
  For example, since high-energy theories are formulated in higher 
dimensions, any low
energy limit will necessarily involve dimensional reduction, and possibly also
compactification \cite{Banks}.
  This turns out to be crucial because as a result of this process the 
low energy,
four-dimensional coupling constants become functions of the radii of 
extra dimensions,
which are often variables.
  One can therefore end up with low-energy effective models in which some of the
``fundamental'' constants of nature are time and/or space-varying quantities.
  There are a number of known examples of such models
\cite{Chodos,Marciano,WuWang,Kiritsis,Alexander}.
  On the other hand, there are recent tentative suggestions of a 
time-variation of the
fine structure constant \cite{Webbetal}, but these  require further 
confirmation.
  It is therefore interesting to study the possible observational 
signatures of such
variations, and in particular to find out how such observational 
signatures constrain
the possible models.
  It turns out to be easier to study this issue by constructing simple 
``toy models''.
  This provides a ``bottom-up'' approach, in which one gives up the 
possibility of
testing particular assumptions from first principles, but instead has 
the possibility
of exploring a larger patch of parameter space.
  This idea goes back at least to Dirac, and had its first detailed 
realization with the
Brans-Dicke model \cite{BransDicke}, which has a varying $G$.
  A number of toy models have recently been constructed to analyse 
possible variations
of the fine structure constant
\cite{Hannestad,Kaplinghat,Bergstrom}, the speed of light
\cite{Mof92,Mof98,BarMag98,AlbreMag99,Bar99,AveMar99,BarMag99,AveMarRoc,AveMar00a} 
and
electric charge \cite{Bekenstein}.

  A somewhat related approach is that of ``quintessence'' (see for example
\cite{PeeblesRatra,Wetterich,Caldwell,Hueyetal}).
  These are essentially models with a time-varying cosmological constant.

  Here we consider the possible observational effects of having a 
universe made up of
different domains, each with a different value of the cosmological constant.
  Such a structure would dramatically influence the future evolution 
of the universe
\cite{Starkman,AvedeCarMar}.
  Recent observations of Type Ia Supernovae up to redshifts of about
$z \sim 1$ \cite{Perlmutter,Riess,Garnavich}, when combined with CMBR 
anisotropy data,
seem to indicate that our patch of the universe is currently 
characterised by the
parameters $\Omega_\Lambda^0 \simeq 0.7$ and
$\Omega_m^0 \simeq 0.3$, implying that the cosmological constant 
become important only
very recently.
  As has been pointed out before, the Supernovae measurements are 
local, and so they can
not be extrapolated all the way to the horizon.
  For example, we could be living in a small, sub-critical bubble, and our local
neighbourhood could have a value of $\Omega_\Lambda^0\simeq0.7$ that 
is uncharacteristic.
  Here we discuss some basic consequences of such a scenario.
  We shall assume that different regions of space have different 
values of the vacuum
energy density, separated by domain walls.
  This can be achieved if there is a scalar field, say $\phi$, which 
within each region
sits in one of a number of possible minima of a time-independent potential.
  The above simplifying assumptions could be relaxed; for example one 
could instead
consider quintessence-type fields. This would introduce quantitative 
differences, but
would not change the basic qualitative results we are discussing.

  In the following section we describe our numerical simulations of 
the evolution of the
domain wall network.
  We then proceed to discuss the basic features of the structure 
formation mechanism for
this scenario in section~\ref{growth}.
  Finally we present our results in section~\ref{results} and discuss 
our conclusions in
section~\ref{conclusions}.

\section{Evolution of the domain walls}
\label{evequations}

  We consider the evolution of a network of domain walls in a k=0
Friedmann-Robertson-Walker universe with line element:

\begin{equation}
  ds^2=a^2(\eta)(d\eta^2-dx^2-dy^2-dz^2),
\label{metric}
\end{equation}
  where $a(\eta)$ is the cosmological expansion factor and $\eta$ is 
the conformal time.
  The dynamics of a scalar filed $\phi$ is determined by the Lagrangian density,

\begin{equation}
  {\cal L}=-{1 \over {4 \pi}}\left({1 \over 2} \phi_{,\alpha} \phi^{,
\alpha} +  V(\phi)\right),
\label{action1}
\end{equation}
  where we will take $V(\phi)$ to be a generic $\phi^4$ potential with 
two degenerate
minima

\begin{equation}
  V(\phi)=V_0\left({\phi^2 \over \phi_0^2}-1\right)^2 + C(\phi).
\label{potential}
\end{equation}
  where $C(\phi)$ smoothly interpolates between $V_-$ at $\phi = - 
\phi_0$ and $V_+$ at
$\phi = + \phi_0$ and is otherwise such that the potential $V(\phi)$ 
has two  minima at
$\pm \phi_0$ which have different energies.
  This obviously  admits domain wall solutions \cite{Vilenkin}.
  The precise form of the function $C(\phi)$ is not important as it 
will not affect
domain walls dynamics if $C(\phi) \ll V_0$ for all possible values of 
the scalar field
$\phi$.
  By varying the action

\begin{equation}
  S=\int d\eta \int d^3x {\sqrt {-g}} {\cal L},
\label{action2}
\end{equation}
  with respect to $\phi$ we obtain the field equation of motion:

\begin{equation}
  {{\partial^2 \phi} \over {\partial \eta^2}} + 2 \frac{\dot a}{a}
{{\partial \phi} \over {\partial \eta}}  - {\nabla}^2 \phi=
-a^{2}{{\partial V} \over {\partial \phi}}.
\label{dynamics1}
\end{equation}
  with

\begin{equation}
  {\nabla}^2={{\partial^2} \over {\partial x^2}} +{{\partial^2} \over
{\partial y^2}} +{{\partial^2} \over {\partial z^2}}.
\label{laplacian1}
\end{equation}

  When making numerical simulations of the evolution of domain wall 
networks (or indeed
other defects) it is also often convenient to modify the equation of 
motion for the
scalar field $\phi$ in such a way that the comoving thickness of the 
walls is fixed in
comoving coordinates.
  This is known as the PRS algorithm \cite{PressRS} and it will not 
significantly affect
the large-scale dynamics of  domain walls.

Hence, we will modify the evolution equation for the scalar field 
$\phi$ according to
the PRS prescription:

\begin{equation}
  {{\partial^2 \phi} \over {\partial \eta^2}} + \beta_1 \frac{\dot a}{a}
{{\partial \phi} \over {\partial \eta}}  - {\nabla}^2 \phi=
-a^{\beta_2}{{\partial V} \over {\partial \phi}}.
\label{dynamics2}
\end{equation}
  where $\beta_1$ and $\beta_2$ are constants.
  We choose $\beta_2=0$ in order  for the walls to have constant 
comoving thickness and
$\beta_1=3$ by requiring  that the momentum conservation law for how 
a wall slows down
in an expanding  universe is maintained \cite{PressRS}.

  We perform two-dimensional simulations of domain wall evolution for 
which $\partial^2
\phi / \partial z^2 =0$.
   These have the advantage of allowing a larger dynamic range and 
better resolution than
tree-dimensional simulations.
  We assume the initial value of $\phi$ to be a random variable 
between $-\phi_0$ and
$\phi_0$ and the  initial value of ${\dot \phi}$ to be equal to zero 
everywhere.
  We normalise the numerical simulations so that $\phi_0=1$.
  We have checked \cite{AveMar00b} that the initial conditions  are 
unimportant (as
expected), because the  domain wall network rapidly approaches a 
scaling solution with
its statistical properties being independent of the initial configuration.

\section{Evolution of cosmological perturbations}
\label{growth}

  As described previously, we shall assume that the universe is made 
up of several
regions (domains), with different vacuum energy densities.
  We will discuss the case where there are two such possible values, but 
it would be easy
to generalise this to a distribution with a continuous range of values.
  Moreover, one assumes that the thin region separating  any two of the domains
considered (domain wall) is not relevant for structure formation.
  This happens if the potential of the field is small enough at the origin.
  Vacuum energy becomes dominant only for recent epochs and so we 
shall be concerned with
the evolution of perturbations only in the matter-dominated era, neglecting the
contribution of the radiation component.
  The average vacuum density is $\rho_\Lambda = (V_+ + V_-)/2$, so that

\begin{equation}
{\overline \Omega}_\Lambda \equiv \rho_\Lambda / \rho_c = {{V_+ + V_-} \over {2
\rho_c}},
\end{equation}
  where $\rho_c$ is the critical density.
  We define $\Delta \Omega_\Lambda$ in a particular domain as

\begin{equation}
\Delta \Omega_\Lambda = \Delta \rho_\Lambda/\rho_c = {{2 V - V_+ - V_-}
\over {2 \rho_c}}
\end{equation}
  where $V$ is the value of the vacuum energy density inside the domain.
  In our case $\Delta \Omega_\Lambda$ can have one of two possible 
values $\Delta
\Omega_\Lambda = \pm |V_+ - V_-|/2 \rho_c$.

In the synchronous gauge, the linear evolution equation for  cold 
dark matter density
perturbations, $\delta_m$, in a flat universe with a non-zero 
cosmological constant can
be written as

\begin{equation}
   \ddot \delta_m + {\cal H} \dot \delta_m - {3 \over 2}{\cal H}^2 \,
\left({{\overline \Omega}_m \delta_m - 2 \Delta
\Omega_\Lambda}\right) = 0,
   \label{one}
\end{equation}
  where the evolution of the scale factor $a$, is governed by the 
Friedmann equation

\begin{equation} {\cal H}^2  = {\cal H}^2_0\left({\overline 
{\Omega}}_m^0 a^{-1} +
(1-{\overline \Omega}_m^0) a^2\right).
\label{two}
\end{equation}
  Here a dot represents a derivative with respect to conformal time, 
the  superscript `0'
means that the quantities are to be evaluated at the present time, 
and ${\cal H} ={\dot
a} / a$.
  Note that the average matter and vacuum energy densities at an 
arbitrary epoch can be
written as

\begin{equation}
{\overline \Omega}_m=\frac{{\overline \Omega}_m^0}{{\overline \Omega}_m^0 +
(1-{\overline \Omega}_m^0) a^3}
  \label{there}
\end{equation}
  and ${\overline \Omega}_\Lambda = 1-{\overline \Omega}_m$, where we 
have also chosen
$a_0 = 1$.
  We start the simulation sufficiently early  (say at a red-shift 
$z_i=10^3$) so that we
do not have to worry about the initial  compensation.
  Because our aim is to investigate the average equation of state of 
each domain we shall
assume the following initial conditions for eq.~(\ref{one}):

\begin{equation}
\delta_m(\eta_i)=0, \qquad {\dot \delta_m}(\eta_i)=0.
\end{equation}
  We will parametrise the density perturbations $\delta_m$ in each domain as a
fluctuation $\Delta \Omega_m = \rho_m \delta_m / \rho_c = \delta_m 
{\overline \Omega}_m$
in the local value of the matter density.
  It is easy to show that the time component

\begin{equation}
\tau_{00} = \frac{3}{8 \pi G} {\cal H}^2 \left(\Delta \Omega_m + 
\Delta \Omega_\Lambda
\right)+\frac{2}{8 \pi G} {\cal H} {\dot \delta},
\label{tau00}
\end{equation}
  of the pseudo-stress-energy tensor, $\tau_{\mu \nu}$, must be compensated on
super-horizon scales (with $\tau_{00} \propto k^4$) \cite{Veeraraghavan}.
  Here, the quantity $-{\dot \delta}/ 3 {\cal H}$ can be interpreted 
as a fractional
variation  in the local expansion rate parametrised by $h$.
  We verified that:

\begin{equation}
\Delta \Omega_\Lambda + \Delta \Omega_m -2\frac{\Delta h}{h}\sim0,
\end{equation}
  where $\Delta h /h = -{\dot \delta}/ 3 {\cal H}$ except near the 
boundary between
different domains as expected from the previous discussion.
  Note that the factor of two in the `Hubble' term of the above 
expression arises because
the Friedmann equation relates ${\cal H}^2$ to the average  matter 
and vacuum energy
densities.

This can be confirmed in Fig.\ 1, which shows the value of the three 
terms above, as
well as of its sum, for a particular simulation.

  The cancellation is almost perfect everywhere except where there are 
domain walls.
  A more detailed study also shows that at the domain walls the area 
where the sum of the
three terms is positive is equal to that where the sum is negative, 
so that these
average out to zero over the whole box.

\section{Results and discussion}
\label{results}

One key assumption of our model is that the  energy scale of
the phase transition which
produced the domain walls is sufficiently low  so that the
domain walls have a negligible  effect on structure formation.
The standard bound of 1 MeV \cite{Vilenkin} obviously
applies here.
 The main consequence of this assumption is that the inhomogeneities 
are not generated
by the domain  walls but are due to the different vacuum  densities 
in different domains.
  The fact that the vacuum densities only become important at late 
times explains why
these inhomogeneities in the local values of the cosmological parameters
$\Omega_\Lambda^0$, $\Omega_m^0$ and $h$ are created only at recent times.
  Different values of $\Omega_\Lambda^0$ and $\Omega_m^0$ lead to 
different linear growth
factors from early times to the present.
  For primordial perturbations in a flat $\Lambda CDM$ model, the quantity

\begin{equation}
g(\Omega_m^0)={{5 \Omega_m^0} \over 2} {\left( 
{\frac{1}{70}+\frac{209 \Omega_m^0}{140}
- \frac{(\Omega_m^0)^2}{140}+  (\Omega_m^0)^{4/7} }\right)}^{-1}
\label{ten}
\end{equation}
  provides a very good fit to the suppression of growth of density perturbations
relative to that of a $\Omega_m^0 = 1$ universe \cite{CPT,Eisenstein}.
  This rescaling was also shown to be valid for generic topological 
defect models for
structure formation on all scales of cosmological interest and for 
any reasonable
combination of the cosmological parameters $h$, $\Omega_m^0$ and 
$\Omega_\Lambda^0$
\cite{AvedeCar}.
  Our model will thus lead to a universe made up of several domains, 
in which the growth
factor $g(\Omega_m^0)$  has different values.
  The result is an inhomogeneous universe today with, for example, the 
abundance of
clusters of galaxies varying from one position to another.
  Hence, we expect the presence of occasional `great walls' separating 
domains with
different values of the cosmological  parameters at high red-shift.
  However, given that in our model the domain walls do not generate 
any cosmologically
relevant fluctuations these `great walls' will simply provide a 
smooth transition between
domains with  different values of the cosmological parameters.
  This can clearly be seen in Figs.\ 1 and 2.

  Given that the length-scale corresponding to these inhomogeneities 
is expected to be
close to the horizon scale (simply due to the dynamics of the domain 
walls), we may not
be able to realize that we live in an inhomogeneous universe.
  This happens because as we look far away we are also looking 
backwards in time and the
universe will get more and more homogeneous as the red-shift of the 
cosmological objects
we are looking at increases (note that $\Delta \Omega_\Lambda = 
\Delta \Omega_\Lambda^0
/ ({\overline \Omega}_m^0(1+z)^3 + (1-{\overline \Omega}_m^0))$ 
decreases very  rapidly
with red-shift).
  In fact, the higher the  red-shift we are looking at the larger is 
the  possibility of
finding a domain with different local values of  the cosmological 
parameters but the
harder is to realize that.
  This is clearly illustrated in Fig.\ 2, which shows  the values of 
$\Omega_\Lambda$,
$\Omega_m$ and $\Delta h/h$  for a particular simulation. One should notice the
different colour schemes  being used for $z=0$ and $z=1$; had we used 
the same colour
scheme for the  earlier redshift, no significant fluctuations would 
be detectable by
visual  inspection.
  This also explains why CMB fluctuations created at  last scattering 
will be completely
negligible.
  Significant CMB fluctuations can only be created at  late times but 
even these are
expected to be small  if we do not live near the edge of our domain.
  However, we know that to be true (if $\Delta \Omega_\Lambda^0$ is 
not too  small)
because otherwise the Universe would look very  anisotropic.

  This makes the observational detection of this effect somewhat non-trivial.
  The best way of doing it should be through the determination  of the 
number density of
objects as a function of redshift in  different directions, assuming 
that one has a
reliable understanding of other possible evolutionary effects.
  Specific examples would be the counting of X-ray or 
Sunyaev-Zel'dovich  galaxy clusters
as a function of the red-shift $z$ such as can be performed by XMM or Planck
\cite{Romer,daSilva}, large-scale velocity flows \cite{Zehavi} or 
gravitational lensing
statistics of extragalactic surveys \cite{Quast}.
  Another possibility  is to look for cosmological anisotropies out to 
$z \sim 1$ with
Supernovae Ia.
  Results from a recent analysis \cite{Kolatt} using a combined sample 
of 79 high and low
red-shift supernovae  are consistent with an homogeneous and 
isotropic universe but  do
not exclude the existence of significant anisotropies on cosmological scales.
Unfortunately, some of their assumptions and results do not apply to our model.
  Such an investigation within the scope of our model is more complex 
because the allowed
variations on the local values of the cosmological parameters are not 
independent and
the spatial distribution of the patches with different values of the 
cosmological
parameters is unknown. A simplified analysis of Supernova
and CMB data constraints was recently performed in
\cite{new1,new2}

\section{Conclusions}
\label{conclusions}

  In this paper we have provided a simple example of a cosmological 
scenario where the
universe becomes inhomogeneous at a very recent epoch, in a way which 
is perfectly
consistent with current observations assuming that we are not very 
close to one of the
boundaries.
  The inhomogeneity arises due to the onset of vacuum energy 
domination, if the value of
the `cosmological' constant is different in different domains.
  This in turn implies that the subsequent dynamics of each patch of 
the universe will be
different, leading to different values of other cosmological 
parameters in each patch,
such as the matter density and the Hubble constant.
  The size of each patch is determined by the dynamics of the domain 
walls but is
expected to be of the order of the horizon (although causality prevents it from
exceeding it).
  This fact makes the detection of the cosmological signatures of this 
kind of model more
difficult given that the contribution of the vacuum energy density 
rapidly becomes
negligible with increasing  red-shift.

\acknowledgements

We are grateful to A. Lasenby, G. Rocha and P. Viana
for useful discussion and comments.

  We thank ``Funda{\c c}\~ao para a Ci\^encia e Tecnologia'' (FCT) for
financial support, and ``Centro de Astrof{\' \i}sica da Universidade do Porto''
(CAUP) for the facilities  provided.
  C.\ M.\ is funded by FCT under  ``Programa PRAXIS XXI'' (grant no.\ PRAXIS
XXI/BPD/11769/97).


\begin{figure}
\vbox{\centerline{
\epsfxsize=1.0\hsize\epsfbox{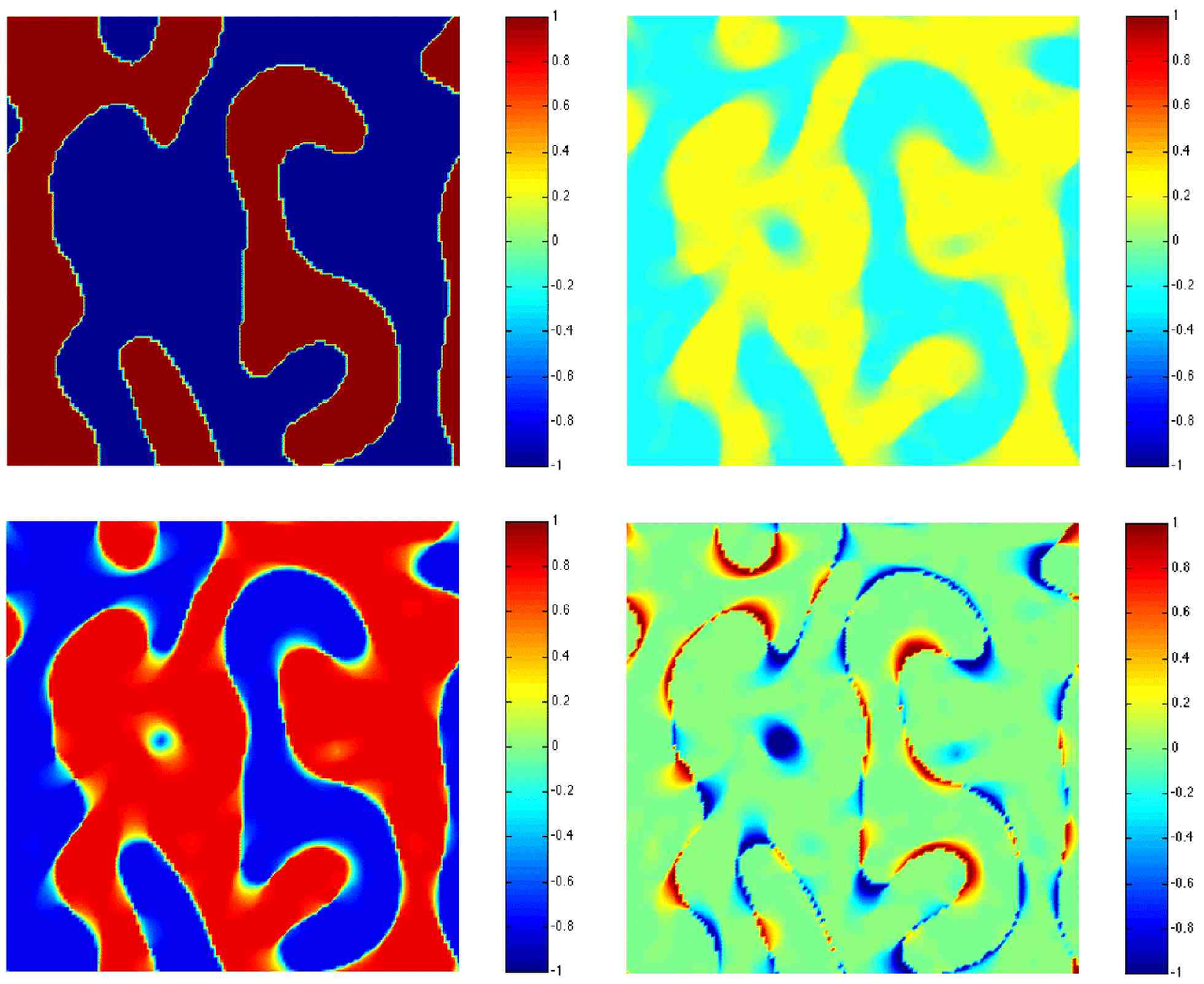}}
\vskip.2in}
\caption{The present day values of $\Delta \Omega_\Lambda$ (top left), $\Delta
\Omega_m$ (top right) and $-2 \Delta h / h$ (bottom left), together 
with their sum
(bottom right) for a particular numerical simulation in units of $|\Delta
\Omega_\Lambda^0|$. The horizon size is about $1/4$ of the box size.}
\label{fig1}
\end{figure}

\begin{figure}
\vbox{\centerline{
\epsfxsize=1.0\hsize\epsfbox{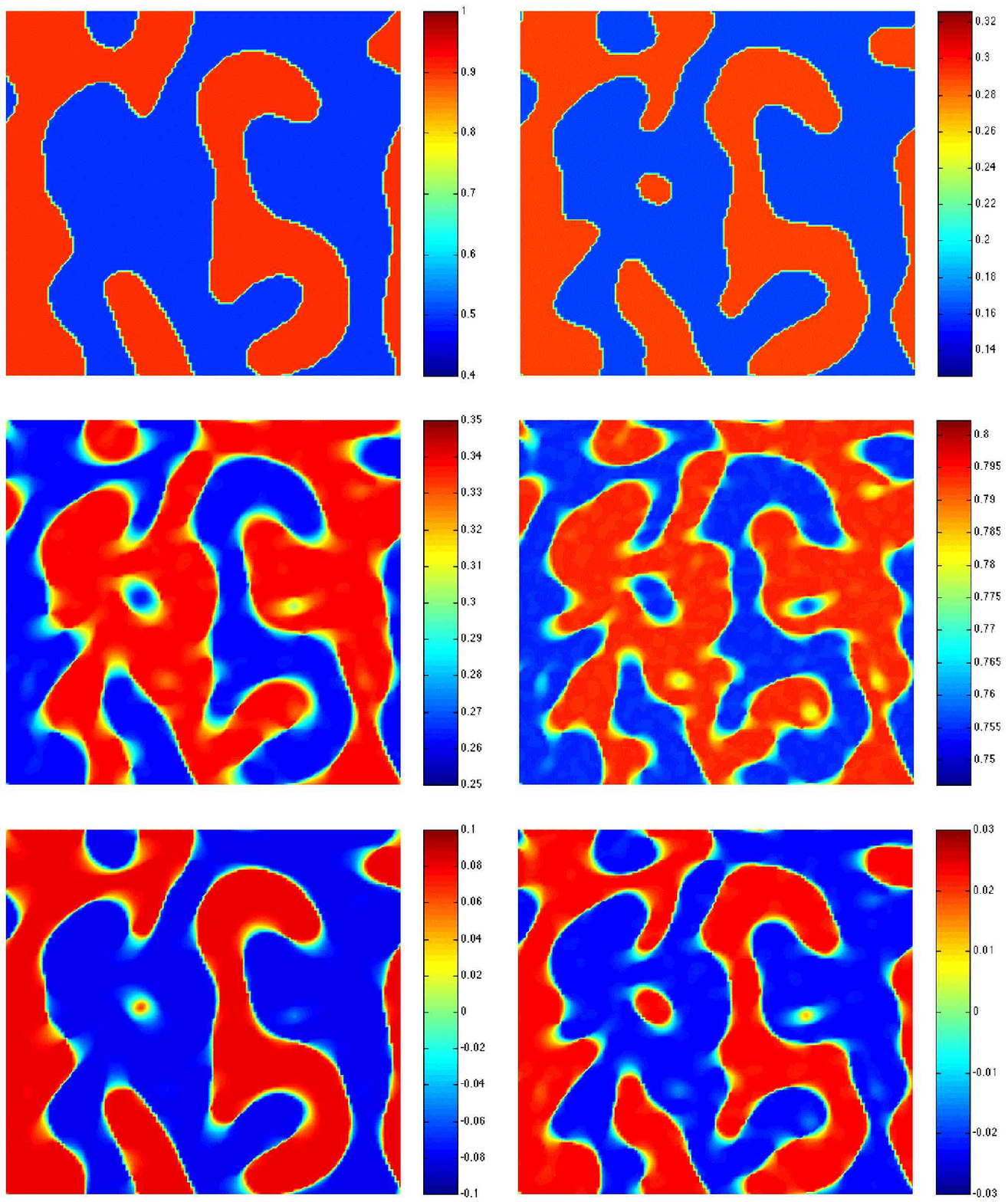}}
\vskip.2in}
\caption{The values of $\Omega_\Lambda = {\overline \Omega}_\Lambda + {\Delta
\Omega}_\Lambda$ (top panels), $\Omega_m  = {\overline \Omega}_m + 
{\Delta \Omega}_m$
(middle panels) and $\Delta h/h$ (bottom panels) for a particular 
simulation box, for
$z=0$ (left) and $z=1$ (right). Notice the different colour scales in 
the left and right
panels. The horizon size at $z=0$ and $z=1$ is respectively about 
$1/4$ and $1/5$ of the
box size.}
\label{fig2}
\end{figure}


\begin{references}
\bibitem{Polchinski}
J.\ Polchinski, {\em String Theory}, Cambridge University Press (1998).
\bibitem{Kolb}
  E.\ W.\ Kolb \& M.\ S.\ Turner, {\it The Early Universe}, Addison-Wesley
(1994).
\bibitem{Banks}
  T.\ Banks, hep-th/9911067 (1999).
\bibitem{Chodos}
  A.\ Chodos \& S.\ Detweiler, {\it Phys.\ Rev.} {\bf D21}, 2167 (1980).
\bibitem{Marciano}
  W.\ J.\ Marciano, {\it Phys.\ Rev.\ Lett.} {\bf 52}, 489 (1984).
\bibitem{WuWang}
  Y.\ S.\ Wu \& Z.\ W.\ Wang, {\it Phys.\ Rev.\ Lett.} {\bf 57}, 1978 (1986).
\bibitem{Kiritsis}
  E.\ Kiritsis, {\it J.H.E.P.} {\bf 10}, 10 (1999).
\bibitem{Alexander}
  S.\ H.\ S.\ Alexander, hep-th/9912037 (1999)
\bibitem{Webbetal}
  J.\ K.\ Webb {\em et al.}, {\it Phys.\ Rev.\ Lett.} {\bf 82}, 884 (1999).
\bibitem{BransDicke}
  C.\ Brans \& R.\ H.\ Dicke, {\it Phys.\ Rev.} {\bf D124}, 925 (1961).
\bibitem{Hannestad}
  S.\ Hannestad, {\it Phys.\ Rev.} {\bf D60}, 023515 (1999).
\bibitem{Kaplinghat}
  M.\ Kaplinghat, R.\ J.\ Scherrer \& M.\ S.\ Turner, {\it Phys.\ Rev.}
{\bf D60}, 023516 (1999).
\bibitem{Bergstrom}
  L.\ Bergstrom, S.\ Iguri \& H.\ Rubinstein, {\it Phys.\ Rev.} {\bf D60},
045005 (1999).
\bibitem{Mof92}
  J.\ W.\ Moffat, {\it Int.\ J.\ Mod.\ Phys.} {\bf D2}, 351 (1992).
\bibitem{Mof98}
  J.\ W.\ Moffat, astro-ph/9811390 (1998).
\bibitem{BarMag98}
  J.\ D.\ Barrow \& J.\ Magueijo, {\it Phys.\ Lett.} {\bf B443}, 104 (1998).
\bibitem{AlbreMag99}
  A.\ Albrecht \& J.\ Magueijo, {\it Phys.\ Rev.} {\bf D59}, 043516 (1999)
\bibitem{Bar99}
  J.\ D.\ Barrow, {\it Phys.\ Rev.} {\bf D59}, 043515 (1999).
\bibitem{AveMar99}
  P.\ P.\ Avelino \& C.\ J.\ A.\ P.\ Martins, {\it Phys.\ Lett.} {\bf B459},
468 (1999).
\bibitem{BarMag99}
  J.\ D.\ Barrow \& J.\ Magueijo, astro-ph/9907354 (1999).
\bibitem{AveMarRoc}
  P.\ P.\ Avelino, C.\ J.\ A.\ P.\ Martins \& G.\ Rocha, {\it Phys.\ Lett.}
{\bf B483}, 210 (2000).
\bibitem{AveMar00a}
  P.\ P.\ Avelino \& C.\ J.\ A.\ P.\ Martins, {\it Phys.\ Rev. Lett.} 
{\bf 85}, 1370
(2000a).
\bibitem{Bekenstein}
  J.\ D.\ Bekenstein, {\it Phys.\ Rev.} {\bf D25}, 1527 (1982).
\bibitem{PeeblesRatra}
  P.\ J.\ Peebles \& B.\ Ratra, {\it ApJ}, {\bf 325}, L17 (1988).
\bibitem{Wetterich}
  C.\ Wetterich, {\it Nucl.\ Phys.} {\bf B302}, 668 (1988).
\bibitem{Caldwell}
  R.\ R.\ Caldwell, R.\ Dave, \& P.\ J.\ Steinhardt, {\it Phys.\ Rev.\ 
Lett.} {\bf 80},
1582 (1998).
\bibitem{Hueyetal}
  G.\ Huey G. {\em et al.}, {\it Phys.\ Rev.} {\bf D59}, 063005 (1999).
\bibitem{Starkman}
  G.\ Starkman, M.\ Trodden \& T.\ Vachaspati, {\it Phys.\ Rev. Lett.} 
{\bf 83}, 1510
(1999).
\bibitem{AvedeCarMar}
  P.\ P.\ Avelino, J.\ P.\ M.\ de Carvalho \& C.\ J.\ A.\ P.\ Martins, 
{\it Phys.\ RevLett.\ } {\bf B501}, 257 (2001).
(2000).
\bibitem{Perlmutter}
  S.\ Perlmutter {\em et al.}, {\it ApJ}, {\bf 517}, 465 (1999).
\bibitem{Riess}
  A.\ G.\ Riess {\em et al.}, {\it Astron. J.}, {\bf 116}, 1009 (1998).
\bibitem{Garnavich}
  P.\ M.\ Garnavich {\em et al.}, {\it ApJ Lett.}, {\bf 493}, L53 (1998).
\bibitem{Vilenkin}
  A.\ Vilenkin \& E.\ P.\ S.\ Shellard, {\em Cosmic Strings and other 
Topological
Defects}, Cambridge University Press (1994).
\bibitem{PressRS}
  W.\ H.\ Press, B.\ S.\ Ryden \& D.\ N.\ Spergel, {\it ApJ}, {\bf 
347}, 590 (1989).
\bibitem{AveMar00b}
  P.\ P.\ Avelino \& C.\ J.\ A.\ P.\ Martins, {\it Phys.\ Rev.} {\bf 
D62}, 103510 (2000).
\bibitem{Veeraraghavan}
  S.\ Veeraraghavan \& A.\ Stebbins, {\it ApJ}, {\bf 365}, 37 (1990).
\bibitem{CPT}
  S.\ M.\ Carroll, W.\ H.\ Press \& E.\ L.\ Turner, {\it Annu.\ Rev.\ 
Astron.\ Astrophys.}
{\bf 30}, 499 (1992).
\bibitem{Eisenstein}
  D.\ J.\ Eisenstein, astro-ph/9709054 (1997).
\bibitem{AvedeCar}
  P.\ P.\ Avelino \& J.\ P.\ M.\ de Carvalho, {\it M.N.R.A.S.} {\bf 
310}, 1170 (1999)
\bibitem{Romer}
  A.\ K.\ Romer, P.\ T.\ P.\ Viana, A.\ R.\ Liddle \& R.\ G.\ Mann, accepted for
publication in {\it ApJ}, (astro-ph/9911499) (2000).
\bibitem{daSilva}
  A.\ C.\ da Silva, D.\ Barbosa, A.\ R.\ Liddle \& P.\ A.\ Thomas, 
astro-ph/9907224
(1999).
\bibitem{Zehavi}
  I.\ Zehavi \& A.\ Dekel, {\it Nature} {\bf 401}, 252 (1999).
\bibitem{Quast}
  R.\ Quast \& P.\ Helbig, {\it Astron.\ Astroph.} {\bf 344}, 721 (1999).
\bibitem{Kolatt}
  T.\ S.\ Kolatt \&O.\ Lahav, astro-ph/0008041 (2000).
\bibitem{new1}
P.\ P.\ Avelino, J.\ P.\ M.\ de Carvalho \& C.\ J.\ A.\ P.\ Martins, 
astro-ph/0103075 (to appear in Phys. Rev. D).
\bibitem{new2}
P.\ P.\ Avelino, A.\ Canavezes, J.\ P.\ M.\ de Carvalho \&
C.\ J.\ A.\ P.\ Martins, astro-ph/0106245.
\end{references}
\end{document}